\documentclass[preprint,12pt,numbers,sort]{elsarticle}
\usepackage{graphicx}
\usepackage{epsfig}
\usepackage{epstopdf}
\usepackage{dcolumn}
\usepackage{bm}
\usepackage{color}
\usepackage{amsmath}
\usepackage{amssymb}
\usepackage[utf8]{inputenc}
\usepackage[T1]{fontenc}
\usepackage{indentfirst}
\usepackage{enumitem}
\usepackage{xspace}
\usepackage[resetlabels]{multibib}
\usepackage[pdfpagelabels]{hyperref}
\hypersetup{breaklinks=true,linktocpage=true,%
bookmarksnumbered=false,%
colorlinks=true,%
linkcolor=blue,%
citecolor=blue,%
urlcolor= blue}
\newcites{sup}{References}

\newcommand{\mytitle}{Nuclear DFT electromagnetic moments in heavy deformed open-shell odd nuclei}

\begin{document}

\begin{frontmatter}

\title{\protect\mytitle}

\author{J. Bonnard$^{a,b}$, J. Dobaczewski$^{a,c}$, G. Danneaux$^d$, M. Kortelainen$^d$}
\address{
$^a$School of Physics, Engineering and Technology, University of York, Heslington, York YO10 5DD, United Kingdom \\
$^b$Universit{\'e} de Lyon, Institut de Physique des 2 Infinis de Lyon, IN2P3-CNRS-UCBL, 4 rue Enrico Fermi, 69622 Villeurbanne, France \\
$^c$Institute of Theoretical Physics, Faculty of Physics, University of Warsaw, ul. Pasteura 5, PL-02-093 Warsaw, Poland \\
$^d$Department of Physics, University of Jyv{\"a}skyl{\"a}, PB 35(YFL) FIN-40014 Jyv{\"a}skyl{\"a}, Finland
}

\date{\today}

\begin{abstract}
Within the nuclear DFT approach, we determined the magnetic dipole
and electric quadrupole moments for paired nuclear states
corresponding to the proton (neutron) quasiparticles blocked in the
$\pi11/2^-$ ($\nu13/2^+$) intruder configurations. We performed
calculations for all deformed open-shell odd nuclei with
$63\leq{Z}\leq82$ and $82\leq{N}\leq126$. Time-reversal symmetry was
broken in the intrinsic reference frame and self-consistent shape and
spin core polarizations were established. We determined spectroscopic
moments of angular-momentum-projected wave functions and compared
them with available experimental data. We obtained good agreement
with data without using effective $g$-factors or effective charges in
the dipole or quadrupole operators, respectively. We also showed that
the intrinsic magnetic dipole moments, or those obtained for conserved
intrinsic time-reversal symmetry, do not represent viable
approximations of the spectroscopic ones.
\end{abstract}

\begin{keyword}
mean field \sep
electromagnetic moments \sep
symmetry restoration \sep
\end{keyword}

\end{frontmatter}

Electromagnetic moments offer a wealth of information for
understanding structure of atomic nuclei. Indeed, while electric
quadrupole moments shed light on shapes of nuclei, and thus probe the
collective degrees of freedom, magnetic dipole moments are more closely
related to single-particle aspects such as shell properties of
valence nucleons~\cite{(Kop58),(Cas90b)}. Over the time, considerable efforts have been
undertaken by experimentalists to develop new techniques
allowing us to access electromagnetic moments of more and more
exotic isotopes~\cite{(Ney03)}. Experimental trends in many isotopic
chains have now been measured, for ground states as well as excited
and isomeric states, see for example recent work in
Ref.~\cite{(Ver22b)}.

\begin{figure}
\begin{center}
\includegraphics[width=\columnwidth]{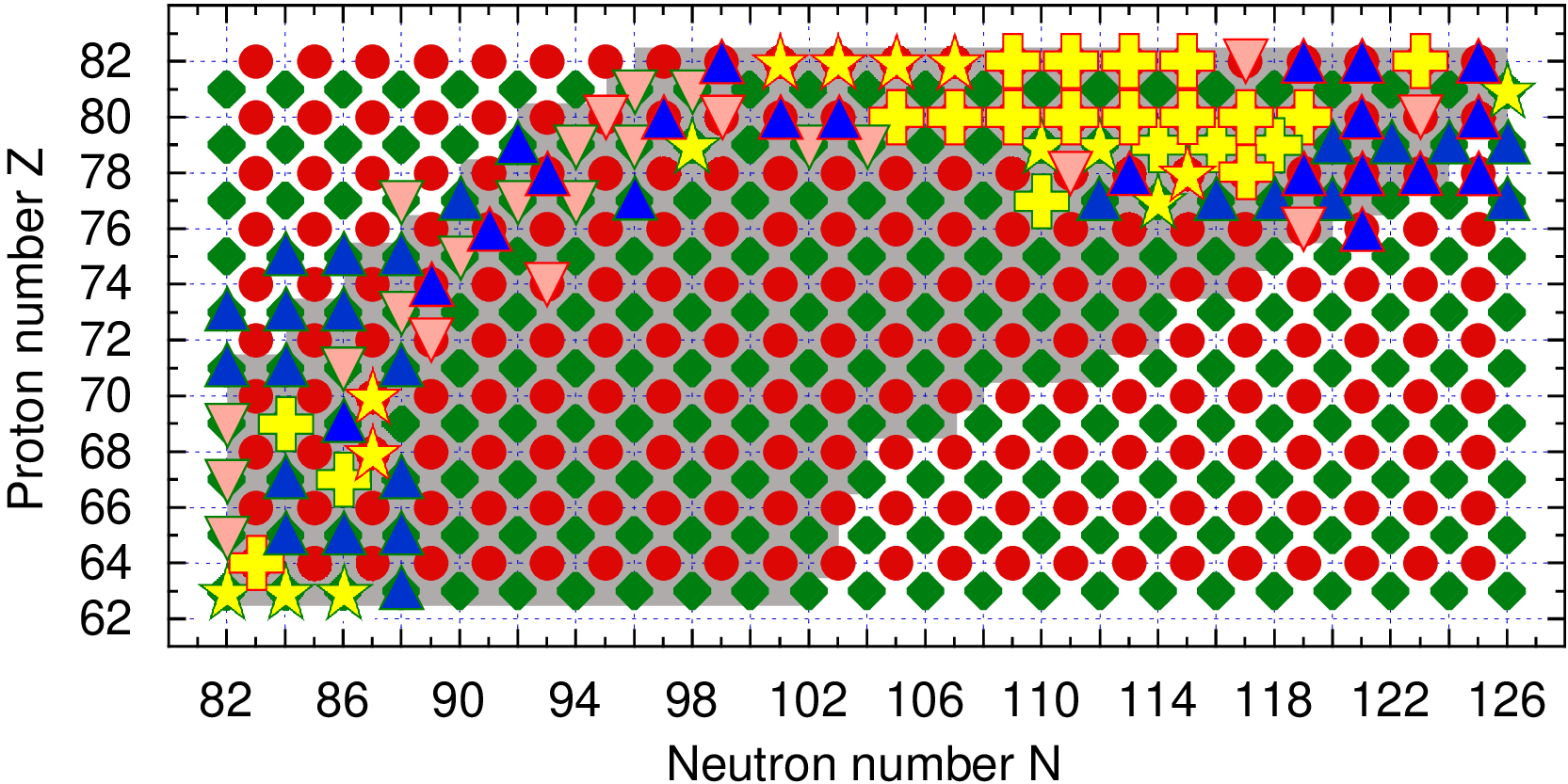}
\end{center}
\caption{Schematic illustration of the set of the odd-$Z$ (diamonds)
and odd-$N$ (circles) nuclei investigated in this work. Stars (plus
signs) represent nuclei where only magnetic dipole moments (magnetic
dipole and electric quadrupole moments) have so far been
measured~\protect\cite{(Sto05d),(Sto14),(Sto16),(Bar20)}. Up
triangles (down triangles) represent nuclei where the $11/2^-$ or
$13/2^+$ states were experimentally identified (tentatively
identified)~\protect\cite{(Kon21)}. Shaded area illustrates the
region of nuclei with known masses~\protect\cite{(Hua21)}.
}
\label{fig:Scheme}
\end{figure}

In this Letter, we present the results of the first systematic
mean-field calculations of spectroscopic electromagnetic moments in
heavy deformed open-shell paired odd nuclei, which offers the
methodology to interpret measurements performed within thousands of
experiments. We focus on the $11/2^-$ and $13/2^+$ isomeric and
ground states of nuclei with proton and neutron numbers of
$63\leq{Z}\leq82$ and $82\leq{N}\leq126$, for which a significant
amount of data is now
available~\cite{(Sto05d),(Sto14),(Sto16),(Bar20)}. In
Fig.~\ref{fig:Scheme}, we show an overview of the considered region
of the Segr\'e chart and indicate nuclei where different types of
experimental data are known.

It appears that no theoretical model of nuclear structure is able
to provide a satisfying unified and global description of electromagnetic
moments. On the one hand, great progress was achieved by \textit{ab initio}
approaches for light species~\cite{Car15}, but their extension to
heavier systems remains challenging because of the necessity to
include extensive many-body correlations~\cite{Hen18}, many-body
currents~\cite{Pas13}, or unmanageably large phase spaces~\cite{(Str22)}.
On the other hand, traditional shell-model
successfully describes experimental electromagnetic
moments~\cite{(Ney03),Pap13,Klo19,(Rod20b)}. Nonetheless, such
calculations often resort to effective $g$-factors or effective charges tuned
to reproduce specific data in limited regions of the chart of
nuclei, so much so that the values obtained in different regions are
not consistent. Finally, nuclear density-functional theory (DFT) has proved
to perform globally well for charge radii and electric quadrupole
moments of even-even nuclei across the nuclear
chart~\cite{Sab07,Del10}. In contrast, however, DFT calculations in
open-shell odd systems still demand further developments to allow
for an overall good description of electromagnetic
moments~\cite{(Bal14d),Bon15,Bor17,Li18,(Per21),(Bar21),(Rys22a),(Bal22)}.

This Letter presents the results of the first systematic calculation
of the spectroscopic magnetic dipole and electric quadrupole moments
that go beyond current state-of-the-art. In a recent
publication~\cite{(Sas22b)}, we have laid out the basics of our DFT
approach and applied it to selected nuclei without pairing
correlations. Here, by including pairing, we port the method's
applicability to all nuclei. This constitutes a major advancement in
the field because, apart from formulating general principles of
symmetry restoration for blocked states~\cite{(Ber12b),(Ave12)} and a
few applications in selected odd nuclei~\cite{(Bal14d),(Bal22)}, no
systematic implementation thereof in heavy odd nuclei existed so far.
Nevertheless, we also note that Refs.~\cite{(Bal14d),(Bal22)} go beyond
our study by including triaxial shapes and generator-coordinate-method
collective correlations.

In addition, we bring about an important new strategy by performing
analyses for specific fixed configurations, irrespective of their
detailed excitation energies above the ground states. This allows us
to follow selected configurations across many nuclei, in function of
numbers of protons or neutrons, and thus to gain better physical
insight into the properties of these configurations. To follow
specific configurations, we implemented the following four-step
procedure.

First, we performed a simple unpaired
calculation of a doubly magic spherical nucleus $^{208}$Pb. In the second step,
by breaking spherical and time-reversal symmetries and using weak
constraints on axial intrinsic quadrupole moment $Q_{20}$ and
angular-momentum projection $I_z$~\cite{(Dob00d)}, we determined weakly prolate and
oblate $^{208}$Pb states with the axial symmetry axis and
single-particle angular momenta aligned along the $z$-axis. At this
point, we could identify the required single-particle wave functions
of states originating from the spherical proton $h_{11/2}$ and
neutron $i_{13/2}$ orbitals that have good values of parity $\pi$ and projection
$\Omega$ on the axial-symmetry axis of $\Omega^\pi=+11/2^-$ and
$\Omega^\pi=+13/2^+$, respectively. These two selected single-particle
states were then fixed and used to tag quasiparticle
states~\cite{(Dob09g),(Ber09c)} in the quasiparticle-blocking
calculations required for the paired odd-$Z$ and odd-$N$ axial nuclei
considered here.

In the third step, we performed tagged quasiparticle-blocking
calculations by fixing the average constant neutron and proton pairing
gaps~\cite{(Dob04d)} at $\Delta_n=\Delta_p=1$\,MeV and constraining
the intrinsic mass quadrupole moments to either $Q_{20}=-10$ or
+10\,b. This allowed us to determine stable starting-point solutions
that, in the fourth-step, were reconverged to the self-consistent
oblate or prolate minima, respectively, by relaxing the constraints
on the quadrupole moments and replacing the fixed pairing gaps
by pairing forces. In addition, next to the closed shells, that is,
for $N=82$, 83, 125, and 126 or $Z=81$ and 82, where the neutron or
proton pairing correlations disappear, we performed calculations with
the paired solutions explicitly switched over to the unpaired
ones~\cite{(Dob21e)}. To follow the same configurations as those
tagged in the quasiparticle-blocking paired solutions, in version
(v3.16n) of code {\sc{hfodd}}~\cite{(Dob23)}, we implemented the tagging
technique also for unpaired states~\cite{data-EuroTal}.

\begin{figure}
\begin{center}
\includegraphics[width=\columnwidth]{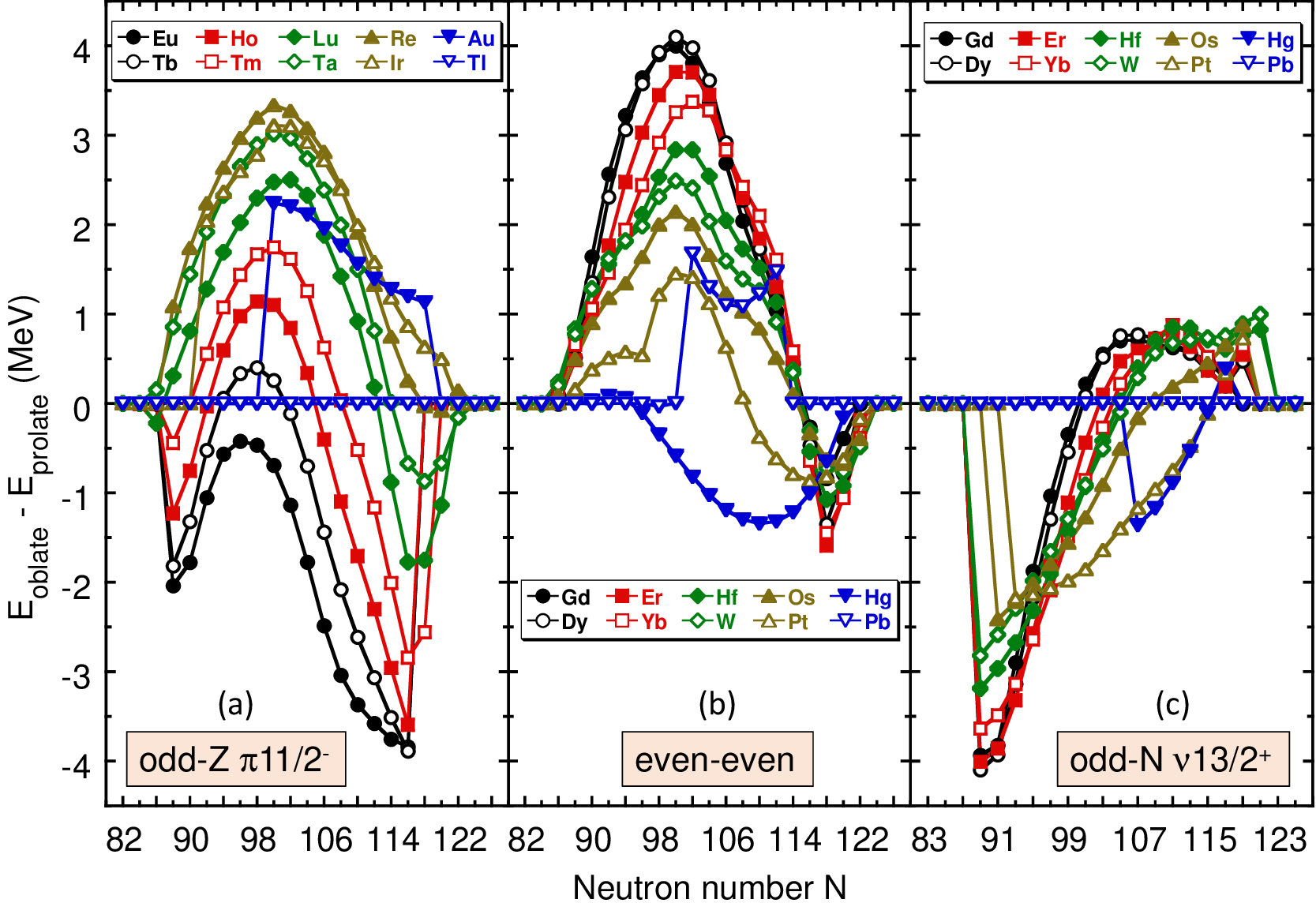}
\end{center}
\caption{Differences of energies at oblate and prolate minima calculated for the
odd-$Z$ $\pi11/2^-$ configurations (a), even-even nuclei (b), and odd-$N$ $\nu13/2^+$ configurations (c).
Lines reverting to the values of $E_{\text{oblate}}-E_{\text{prolate}}=0$ denote cases where
either oblate or prolate higher-energy minimum disappears.
}
\label{fig:Eprol-Eobl}
\end{figure}

As it turned out, for the majority of the studied nuclei, both oblate
and prolate minima were obtained. However, as shown in
Fig.~\ref{fig:Eprol-Eobl}, the oblate-prolate energy differences
obtained in the odd-$Z$ and odd-$N$ nuclei were very different from
one another, and also both were very different from those in
neighboring even-even nuclei. It appears that without a fully
self-consistent calculation, in an open-shell nucleus one cannot
{\it a priori} guess how an odd particle would polarize the even system.

Finally, we determined the standard spectroscopic
moments~\cite{(Rin80b)} by performing the angular-momentum projection
(AMP)~\cite{(She21)} of the obtained intrinsic paired and unpaired
states. This was done using version (v3.16n) of code
{\sc{hfodd}}~\cite{(Dob21e),(Dob23)}, with the Pfaffian method used
to determine overlaps of blocked paired
states~\cite{(Rob09),(Ber12b)}. We also tested the effects of the
particle-number projection (PNP)~\cite{(She21),(Dob21e)}, performed
together with the AMP. This turned out to give the magnetic dipole
and electric quadrupole moments within about 1\% of the AMP-only
values~\cite{supp-EuroTal}. Therefore, the AMP+PNP calculations,
which take about two orders of magnitude more CPU time than those
performed for the AMP-only, can be safely avoided.

In this Letter, we considered the Skyrme functional
UNEDF1~\cite{(Kor12b)} with the pairing correlations treated within
the standard two-basis method~\cite{(Gal94),(Sch12c)} for the single-particle
states cut at 60\,MeV. The neutron and
proton pairing strengths were increased by 20\% with respect to the
original UNEDF1 parameter adjustment. Tests have shown that such an
increase compensates for the effects of the Lipkin-Nogami approximate
PNP~\cite{(She21)}, which was not used in this work. An essential
element of our approach is the mean-field time-reversal-symmetry
breaking induced by the spin-spin interaction implemented in terms of
the isovector Landau parameter $g_0'$. We used the value of
$g_0'=1.7(4)$ adjusted in Ref.~\cite{(Sas22b)}, see also the Supplemental
Material~\protect\cite{supp-EuroTal}, and there were no
other parameters adjusted in the present work.

\begin{figure}
\begin{center}
\includegraphics[width=\columnwidth]{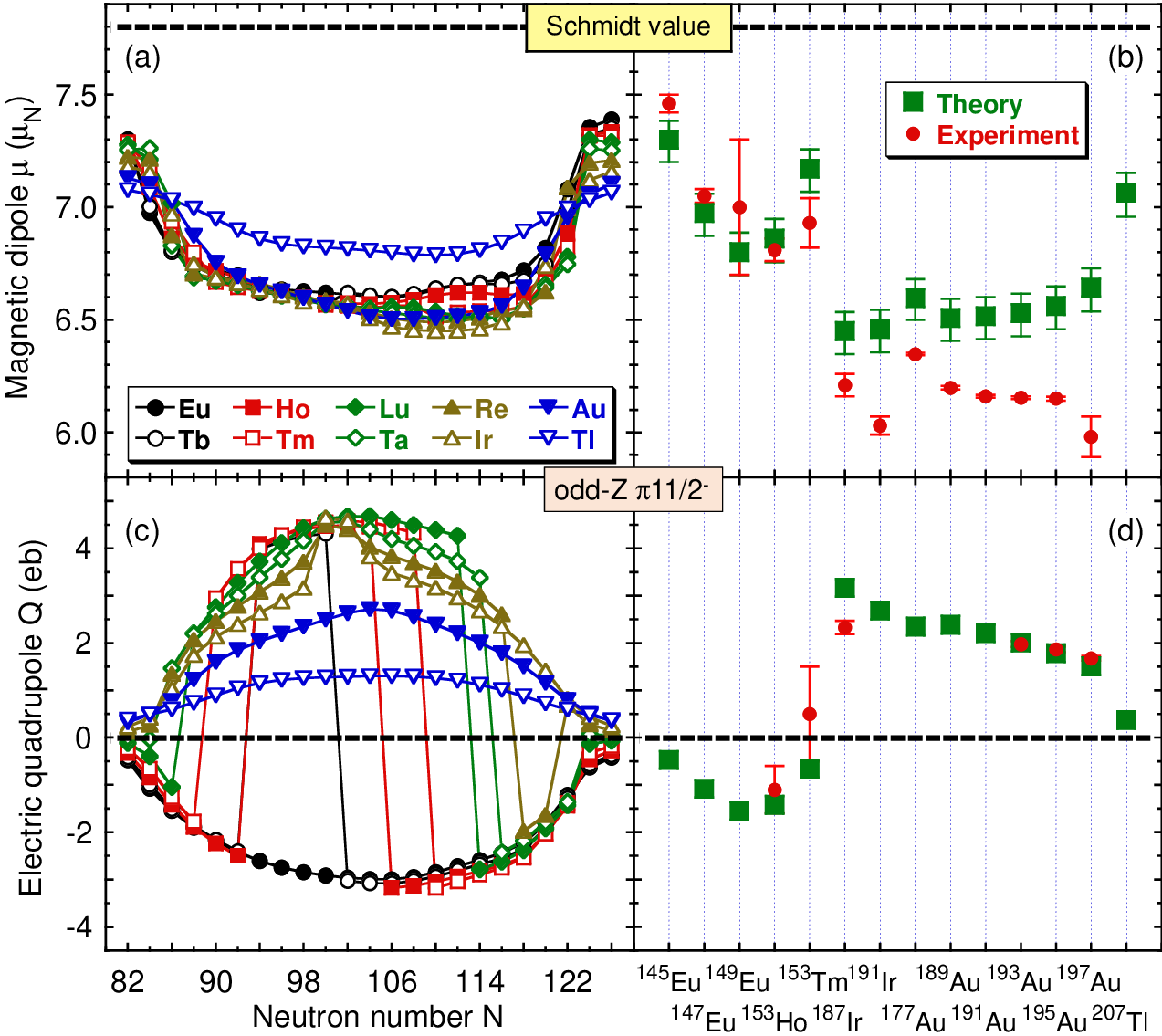}
\end{center}
\caption{Calculated magnetic dipole $\mu$ (a) and electric quadrupole $Q$ (c) moments
of the $\pi11/2^-$ configurations in odd-$Z$ nuclei compared in panels (b) and (d) with the available
experimental data~\protect\cite{(Sto05d),(Sto14),(Sto16),(Bar20)}. Theoretical error bars
correspond to the uncertainty of the isovector Landau parameter $g_0'=1.7(4)$, see details presented
in Refs.~\cite{(Sas22b),supp-EuroTal}.
}
\label{fig:EuroTal}
\end{figure}

\begin{figure}
\begin{center}
\includegraphics[width=\columnwidth]{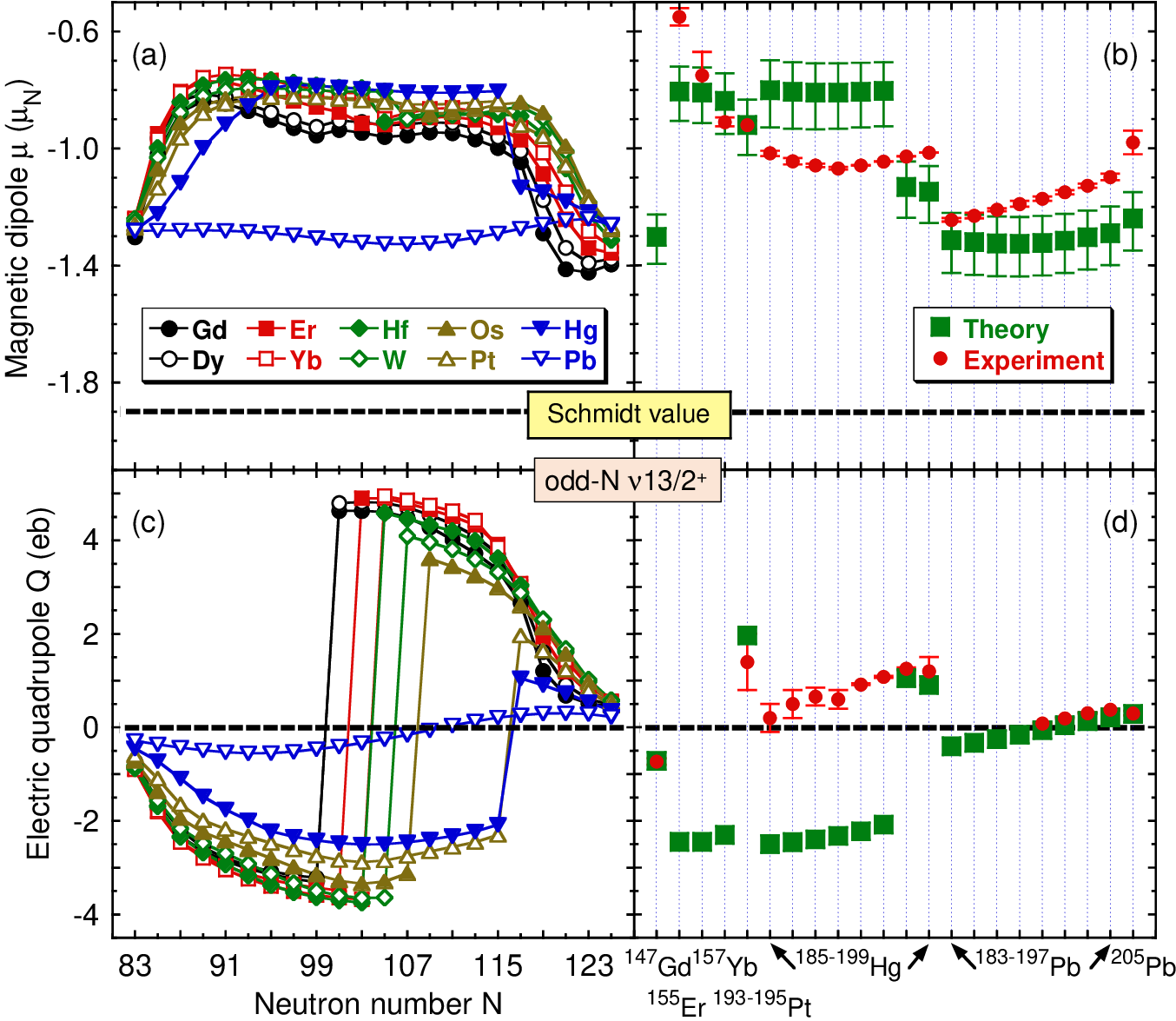}
\end{center}
\caption{Same as in Fig.~\protect\ref{fig:EuroTal} but for the $\nu13/2^+$ configurations in odd-$N$ nuclei.
}
\label{fig:GadLead}
\end{figure}

Figures~\ref{fig:EuroTal} and~\ref{fig:GadLead} summarize principal
results of our study obtained for the $\pi11/2^-$ configurations in
odd-$Z$ nuclei and $\nu13/2^+$ configurations in odd-$N$ nuclei~\cite{supp-EuroTal}.
Across the entire set of nuclei studied here, the corresponding two
deformed single-particle states are characterized by the Nilsson
labels~\footnote{We determined the Nilsson label of a given
self-consistent single-particle state by finding the largest
amplitude of its expansion onto the deformed axial harmonic-oscillator
basis~\protect\cite{(Dob00d)}.} [505]11/2 and [606]13/2, respectively.
Within the studied proton and neutron shell, with increasing numbers
of protons or neutrons, the occupation numbers of states [505]11/2 and
[606]13/2 are swept between zero and one. This allows us to gain a
unique insight into the polarization properties of these two states in
the function of the numbers of particles.

The electric quadrupole moments $Q$, shown in
Figs.~\ref{fig:EuroTal}(c) and \ref{fig:GadLead}(c), present clear
patterns of the particle-number dependence, which has approximately a
parabolic shape both on the prolate and oblate side. Small values of
$Q$ obtained near closed shells follow the well-known scheme dictated
by the deformation-splitting of spherical orbitals. Indeed, there the
high-$\Omega$ magnetic substates have the highest (lowest) energies
on the prolate (oblate) side, and thus one obtains prolate (oblate)
shapes when these orbitals are occupied by holes (particles). What is
much less obvious, and what is very clearly born out in our results,
is the fact that the same signs of the quadrupole moments are
obtained when the configurations are followed away from the closed
shells. This has very important consequences, namely, a configuration
that begins as a hole (particle) state at the lower end of the
particle numbers, somewhere between the magic numbers {\em must} change the
sign of the quadrupole moment. This is so because at the higher end,
it terminates as a particle (hole) state. This rule is strictly
obeyed in all our results, even if in the middle of the shell the
blocked quasiparticles can be below or above the Fermi surface.

Patterns of the particle-number dependence of magnetic dipole moments
$\mu$, Figs.~\ref{fig:EuroTal}(a) and \ref{fig:GadLead}(a), are
different. Near the neutron closed shells, where the quadrupole
moments are small, their departures from the single-particle Schmidt limits~\cite{(Sch37)} are
dictated by the spin polarizations induced by the time-odd mean
fields~\cite{(Sas22b)}. Detailed values of these departures can depend
relatively strongly ($\pi11/2^-$ near $N=126$) or relatively weakly
($\nu13/2^+$ near $N=82$) on the numbers of protons in the
open shell. With particle numbers moving away from the neutron closed
shells, the magnetic dipole moments undergo rather sudden changes and then,
in the middle of the shell, their values stabilize. This pattern does
not follow the gradual dependence of the quadrupole moments on
particle numbers. Moreover, in open-shell nuclei, the magnetic dipole
moments appear to be independent of whether the nucleus is
oblate or prolate. In addition, for relatively weakly deformed
isotopes of Pb and Tl, the dependence of the magnetic dipole moments on the
number of neutrons is relatively weak.

\begin{figure}
\begin{center}
\includegraphics[width=\columnwidth]{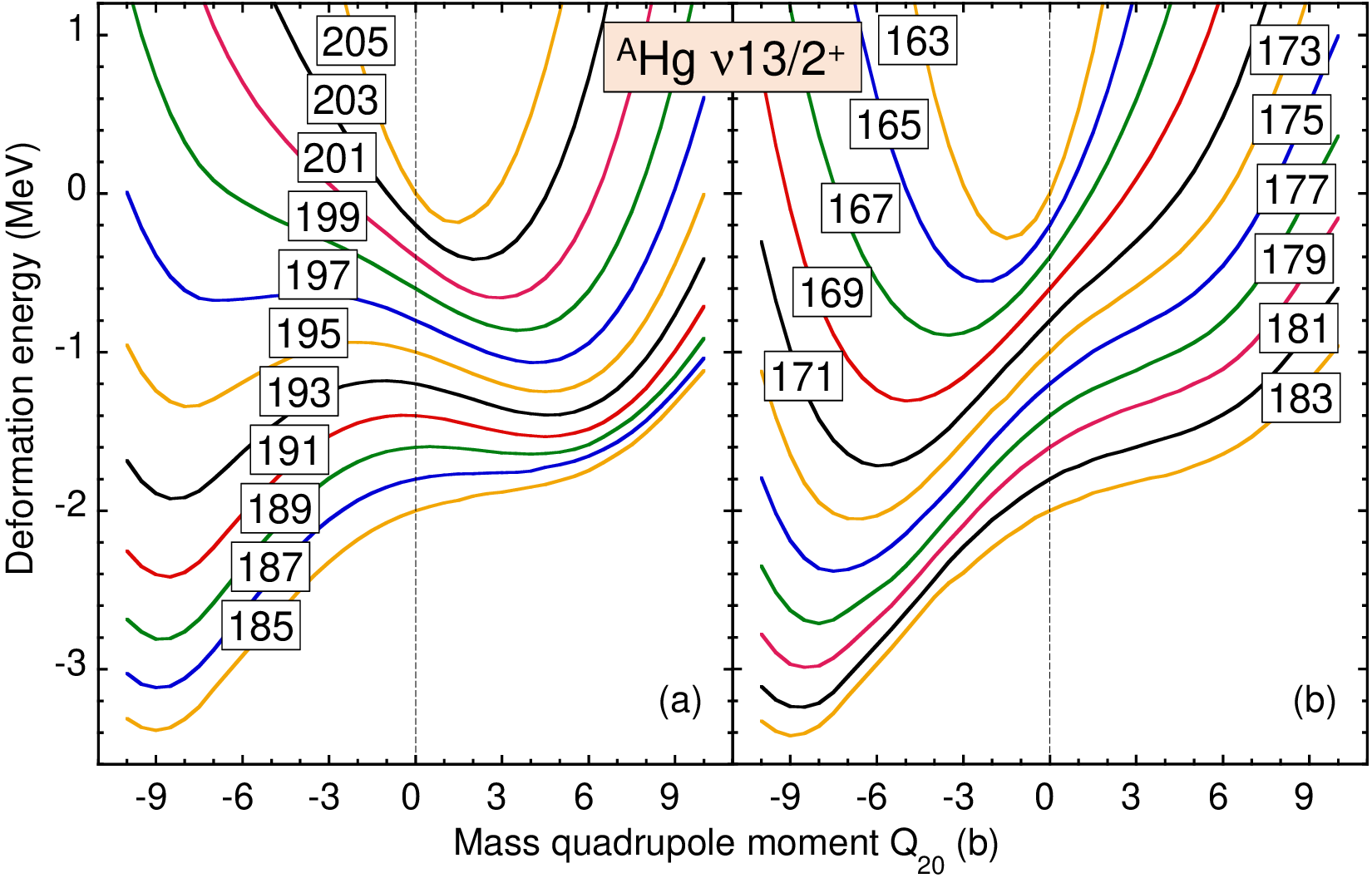}
\end{center}
\caption{Deformation energies, $E(Q_{20})-E(0)$, of the $\nu13/2^+$ isomers in the Hg isotopes as functions of
the intrinsic mass quadrupole moments $Q_{20}$. For better readability, at $A\geq185$ (a) and $A\leq183$ (b),
the curves are shifted by $(A-205)\times0.1$\,MeV and  $(163-A)\times0.1$\,MeV, respectively. Dashed lines mark
points at $Q_{20}=0$.
}
\label{fig:Hg-PES}
\end{figure}

Apart from the case of light Hg isotopes, electric quadrupole moments
agree with data spectacularly well, Figs.~\ref{fig:EuroTal}(d)
and~\ref{fig:GadLead}(d). Also those of the two heaviest measured Hg
isotopes are reproduced perfectly. However, in the calculations,
lighter Hg isotopes undergo a transition to oblate shapes, which is not
born out in the data. As discussed above, such a transition to oblate
shapes must occur somewhere in the isotopic chain because the oblate shape
is mandated by the particle character of the orbital $\nu13/2^+$ near
$N=82$. As clearly visible in the results shown in
Fig.~\ref{fig:Eprol-Eobl}(c) and~\ref{fig:Hg-PES}, not only the
transition to oblate shapes occurs at $N=115$, but below $N=107$, the
prolate minimum disappears entirely.

It appears that the same transition to oblate shapes in light Hg
isotopes creates a deviation from the smooth trend observed in the
measured magnetic dipole moments, Figs.~\ref{fig:EuroTal}(b)
and~\ref{fig:GadLead}(b). Otherwise, the calculated magnetic dipole moments
of heavier nuclei seem to call for stronger time-odd polarization
effects, whereas those of lighter nuclei appear to be about
right~\footnote{After performing the calculations, we learned that the
magnetic dipole moment of $^{207}$Tl has already been measured.
The obtained experimental value was not published yet.}.
Such detailed comparison with data may indicate that
more structured interactions in the time-odd mean-field sector
should possibly be considered beyond the simple spin-spin interaction
employed here.

The magnetic dipole moment of $^{147}$Gd is an exception. In this
nucleus, the calculated value of  $-1.30\,\mu_N$ disagrees with the
two measured values of $-0.24(7)\,\mu_N$~\cite{(Hau79),(Hau82)} and
$+0.49(2)\,\mu_N$~\cite{(Daf87)}, which, in addition, are strongly
incompatible with one another. A possible mixing of the pure
$i_{13/2}$ configuration with the $3^-\otimes f_{7/2}$ contribution
was discussed in the literature~\cite{(Daf87)} and in the future work
it will also be considered within the present DFT approach. However,
such a mixing should neither strongly affect the magnetic dipole
moments of neighboring nuclei nor spoil the excellent agreement with
the value of the quadrupole moment measured in $^{147}$Gd.

\begin{figure}
\begin{center}
\includegraphics[width=\columnwidth]{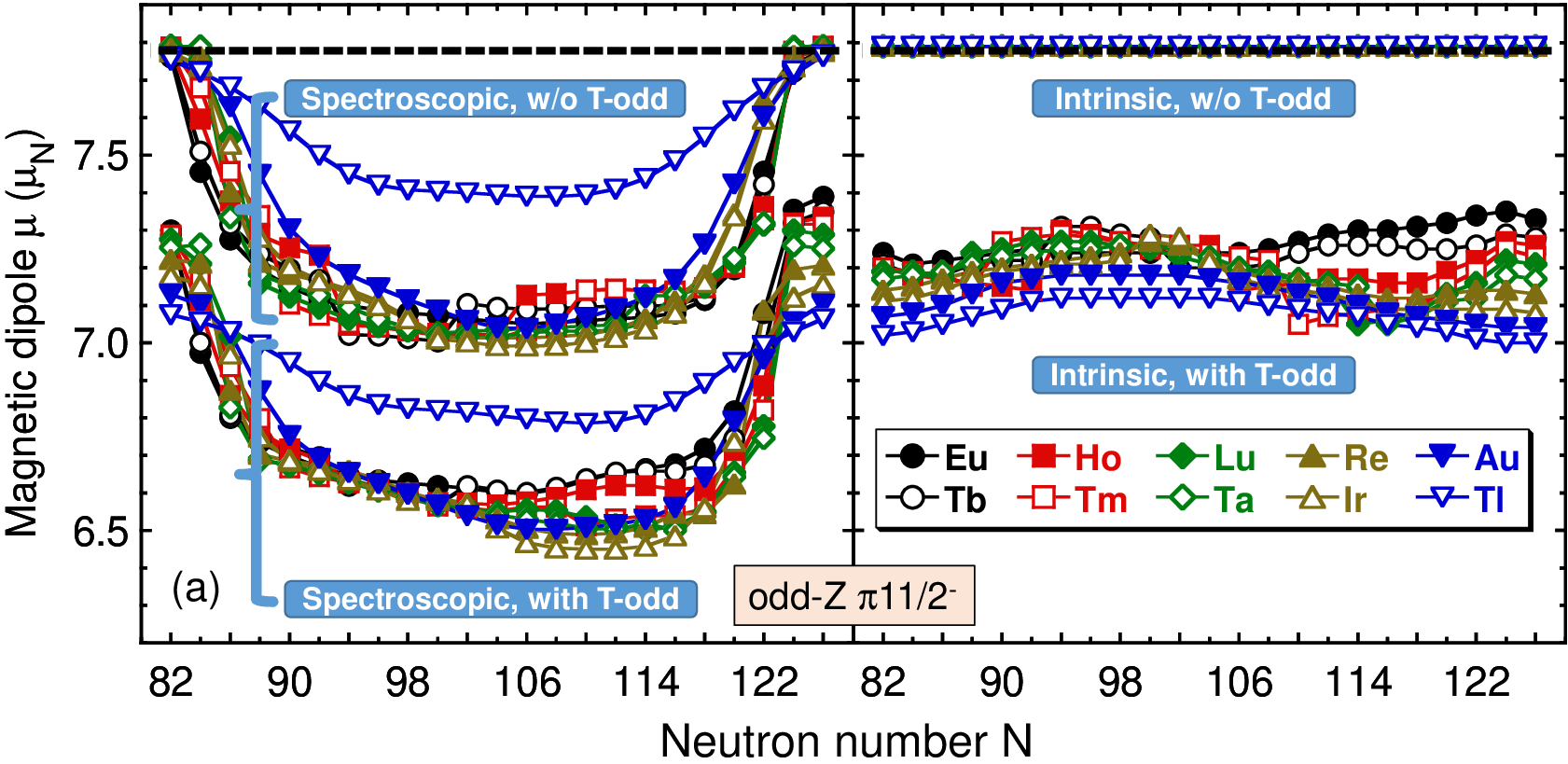}
\end{center}
\caption{Spectroscopic (a) and intrinsic (b) magnetic dipole moments
calculated for the $\pi11/2^-$ configurations in odd-$Z$ nuclei with
and without the time-odd mean fields.
}
\label{fig:intrinsic-mag-h11}
\end{figure}

We conclude our presentation with a comparison of the spectroscopic and
intrinsic magnetic dipole moments determined with and without the
time-odd mean fields, Fig.~\ref{fig:intrinsic-mag-h11}. First, we note
that without the time-odd mean fields, the calculated intrinsic magnetic dipole
moments of all nuclei are exactly equal to the Schmidt
limit~\cite{(Sch37)}. This is a simple consequence of the fact that
in the intrinsic reference frame, the high-spin intruder
configurations studied here have both the orbital and spin angular
momenta always maximally aligned along the symmetry axis. This
property turns out to be entirely independent of the intrinsic
deformation.

With the time-odd mean fields included, the intrinsic magnetic dipole
moments move away from the Schmidt limit and become particle-number
dependent. However, their values stay entirely uncorrelated with the
nuclear deformation. The effect of the deformation only appears when
the spectroscopic magnetic dipole moments are evaluated,
Fig.~\ref{fig:intrinsic-mag-h11}(a). As can be demonstrated
through the particle-core-coupling analysis~\cite{(Ver22b),(Bon23)},
the deformation effects can be traced back to the coupling of the
single-particle wave functions with the $2^+$, $4^+$, $6^+$,{\ldots}
states of the core. These effects turn out to be invisible in the intrinsic
reference frame and require the determination of the spectroscopic moments.

The inclusion of the time-odd mean fields breaks the signature
symmetry of the core and results in the appearance of the $1^+$,
$3^+$, $5^+$,{\ldots} states. The coupling of the single-particle
wave functions with the core's odd-spin states shifts the magnetic
moments away from the Schmidt limit. The magnitude of this shift does
not seem to be correlated with the occupations of the spin-orbit
partners, which vary within the single-particle shells. This is in
contrast to the traditional shell-model interpretation, which is
based on the so-called first-order correction to the magnetic
moments~\cite{(Cas90b)}. In the nuclear DFT, the effects of the
time-odd mean fields rather seem to proceed through a democratic
(collective) spin polarization of the core.

Since the principal part of the magnetic dipole moment is given by
the contribution of the odd particle, the collectivity of the core
contribution does not result in a meaningful approximation of the
spectroscopic magnetic dipole moment by the intrinsic one. A
microscopic evaluation of the spectroscopic magnetic dipole moments is thus
mandatory~\footnote{To a lesser degree, the same observation also holds for
the spectroscopic electric quadrupole moments, which, at large
deformations, can be precisely approximated by products of the
Clebsh-Gordan coefficient squared and intrinsic electric quadrupole moments.
This very well known fact is illustrated in the Supplemental
Material~\protect\cite{supp-EuroTal}, where we also show that at
small deformations, the relative differences can reach 30\%.}.
Moreover, the core contribution to the magnetic dipole moment does
not seem to be amenable to simple modelization in terms of a smooth
function of $Z/A$, as is the case for the collective
rotation, which generates a collective current flow~\cite{(Boh75a)}.
Altogether, both the angular-momentum symmetry restoration and
time-reversal symmetry breaking allowing for inclusion of the
time-odd mean fields appear to be essential elements of describing
the nuclear magnetic dipole moments.

In conclusion, we presented the results of the first nuclear DFT
determination of the magnetic dipole and electric quadrupole moments
established in 450 heavy open-shell deformed and paired odd nuclei.
We showed that by following fixed single-particle configurations, one
could pin down generic properties of the calculated moments and
analyze their dependence on the proton and neutron numbers. Apart
from pronounced differences that we examined in the Hg isotopes and
$^{147}$Gd, we obtained good agreement with known experimental data.
For that, no effective $g$-factors or effective charges were needed.
We demonstrated an essential role played by the time-odd core
polarization and angular-momentum symmetry restoration. We compared
the calculated spectroscopic and intrinsic magnetic dipole moments
and showed that the former cannot be meaningfully approximated by the
latter.

Our work motivates further experimental and theoretical studies of
nuclear electromagnetic moments. In the experiment, depending on the
lifetimes of states and availability of beams, moments of many more
states 11/2$^-$ and 13/2$^+$ can probably be measured~\cite{(Bar20)}
and compared with our model predictions. This can further inform the
nuclear-DFT calculations and allow for needed improvements and
extensions to exotic moments. In particular, the role of the coupling
between single-particle states and negative-parity core states should
be studied and evaluated globally. In addition, multi-reference
configuration interaction of different single-particle configurations
and/or different $K$-components related to triaxiality should be
analyzed. But first and foremost, calculations pioneered in this work
can become the basis of determining electromagnetic moments across
the entire Segr\'e chart for all odd nuclei and spins and parities.
Work on this extensive goal is now in progress.

\bigskip\noindent
We thank A.N.\ Andreyev, A.E.\ Barzakh, and I.D.\ Moore for useful discussions.
This work was partially supported by the STFC Grant
Nos.~ST/P003885/1 and~ST/V001035/1, by the Polish National
Science Centre under Contract No.~2018/31/B/ST2/02220,
by a Leverhulme Trust Research Project Grant,
and by the Academy of Finland under the Academy Project No.~339243.
We acknowledge the CSC-IT Center for Science
Ltd., Finland, for the allocation of computational resources.
This project was partly undertaken on the Viking Cluster,
which is a high performance compute facility provided by the
University of York. We are grateful for computational support
from the University of York High Performance Computing
service, Viking and the Research Computing team.

\bibliographystyle{elsarticle-num}

\end{document}